# Angle dependent localized surface plasmon resonance from silver nanoparticles embedded in $SiO_2$ thin film

R. K. Bommali[1a)], D. P. Mahapatra[1], H. Gupta[3], Puspendu Guha[1,2], D. Topwal[1,2], G.Vijaya Prakash[3], S. Ghosh[3] and P. Srivastava[3]

[1]*Institute of Physics, Sachivalaya Marg, 751005, Bhubaneswar, India*

[2]*Homi Bhabha National Institute, Training School Complex, Anushakti Nagar, Mumbai 400085, India*

[3]*Department of Physics, Indian Institute of Technology Delhi, Hauz Khas, 110016, New Delhi, India.*

Near surface silver nanoparticles embedded in silicon oxide were obtained by 40 keV silver negative ion implantation without the requirement of an annealing step. Ion beam induced local heating within the film leads to an exo-diffusion of the silver ions towards the film surface resulting in the protrusion of larger nanoparticles. Cross-sectional transmission electron microscopy (XTEM) reveals the presence of poly-disperse nanoparticles (NPs), ranging between 2 nm-20 nm, at different depths of the $SiO_2$ film. The normal incidence reflectance spectrum shows a double kink feature in the vicinity of 400nm, indicating a strong localized surface plasmon resonance (LSPR) from the embedded NPs. However, due to overlap of the bilayer interference and LSPR, the related features are difficult to separate. The ambiguity in associating the correct kink with the LSPR related absorption is cleared with the use of transfer matrix simulations in combination with an effective medium approximation. The simulations are further verified with angle dependent reflectance measurements. Additionally, transfer matrix simulation is also used to calculate the electric field intensity profile through the depth of the film, wherein an enhanced electric field intensity is predicted at the surface of the implanted films.

The phenomena of localized surface plasmon resonance (LSPR) in noble metal nanoparticles has wide variety of application ranging from high resolution optical imaging beyond the diffraction limit[1], surface enhanced raman spectroscopy (SERS)[2], enhanced light trapping in solar cells[3], luminescence yield[4,5], and photo-catalytic activity in $TiO_2$[6].

Embedded nanoparticles (NPs) have been recently demonstrated to be of potential interest as SERS substrates, wherein they offer multiple advantages like smooth surface, chemical stability, reusability and tunability of the LSPR[7]. The embedded nanostructures can be obtained with great reproducibility, at desired depths and high throughput by low energy ion-implantation as demonstrated by several recent reports[8,9,10]. Implantation of ions at a desired depth within an optically transparent film, has an additional benefit that a strong coupling of the bilayer interference in the thin film and LSPR of the NP can lead to a further enhancement of the local electromagnetic fields. In principle, with the choice of transparent hosts like $SiO_2$, $Si_3N_4$ and $TiO_2$ the LSPR in Ag NPs can be shifted to 400nm, 480nm and 600nm respectively[11]. Further, the increased dielectric constant of the surrounding medium has been predicted to enhance the confinement of the LSPR modes

[a)]Author Electronic mail: ravibommali06@gmail.com

in the NPs leading to enhanced LSPR related absorption[11]. The best approach to identify and study the LSPR from such thin films on opaque substrates (c-Si) is by recording the optical reflectance from such samples. However, there has been intrinsic difficulty in resolving the LSPR accurately. The reflectance signal obtained from a thin film containing embedded NPs is composed of overlapped features resulting from thin film interference effect as well as the LSPR related absorption from the NPs, which are difficult to resolve accurately. Previously, reported work[7] in the literature predominantly employs a method of reflectance contrast, requiring the subtraction of pristine film reflectance from the implanted film reflectance. In the present work we find that the combination of the effective medium approximation and the transfer matrix theory can be used to accurately simulate the reflectance spectra from films containing embedded NPs, thereby allowing one to extract optical response of the LSPR from the embedded NPs non-destructively.

Si (100) wafer was oxidized in a box furnace in ambient atmosphere for the growth of the $SiO_2$ thin film on the Si wafer. Negative silver ion implantation has been carried out at the multi cathode source of negative ions by cesium sputtering (MC-SNICS) ion implanter, inter-university accelerator center New Delhi, with energy tuned at 40 keV at beam current density of 1.7 μA/cm$^2$ and an implantation dosage of $3.5\times10^{16}$ ions/cm$^2$. X-ray diffraction (XRD) measurements have been performed on a Philips X'Pert Pro (Model PW 3040) X-ray diffractometer in glancing incidence (1°) mode, after the required alignment steps. UV-visible reflectance measurements have been carried out using UV3600: Shimadzu. Cross-sectional TEM (XTEM) of the as-implanted sample has been carried out on the JEOL JEM 2010 instrument at 200kV accelerating voltage. SEM was carried out on FESEM Carl Zeiss Neon 40 model at 20keV electron energy. Angle dependent reflectance was carried out on Ocean Optics RSS-VA variable angle reflectance measurement attachment, which allowed for measurement reflected intensities for angle of reflection ranging between 15°-45°. DH2000 UV-Vis light source and Maya 2000 Pro spectrometer of Ocean Optics were used for the incident light and detection of the reflected light, respectively. The spectrometer is interfaced with a computer to facilitate data acquisition. The calculations were performed in Matlab.

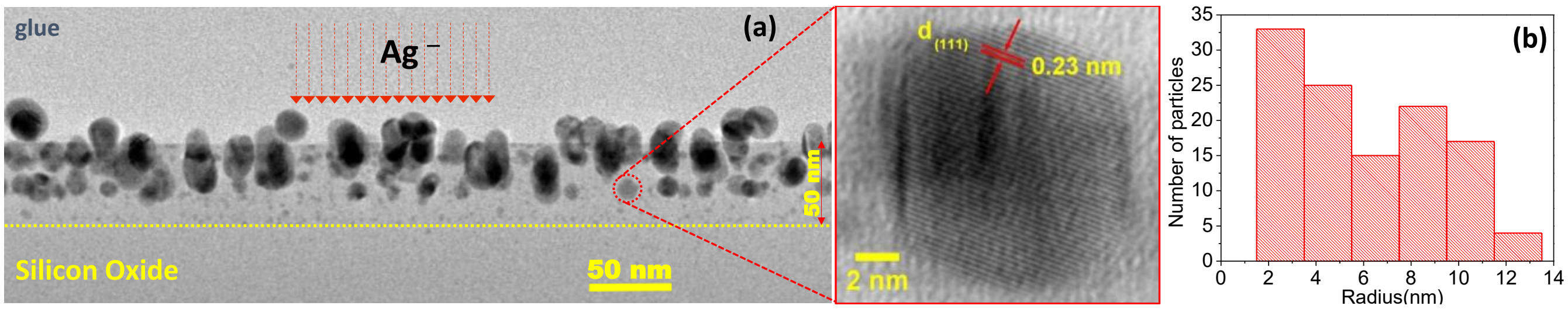

FIG.1. (a) The-Cross sectional TEM image of the silver implanted $SiO_2$ film/Si(100). The red arrows show the direction of silver ion implantation. Towards the right, high resolution image shows single crystalline nature of the formed Ag nanoparticles. (b) The corresponding Size distribution of the Ag nanoparticles.

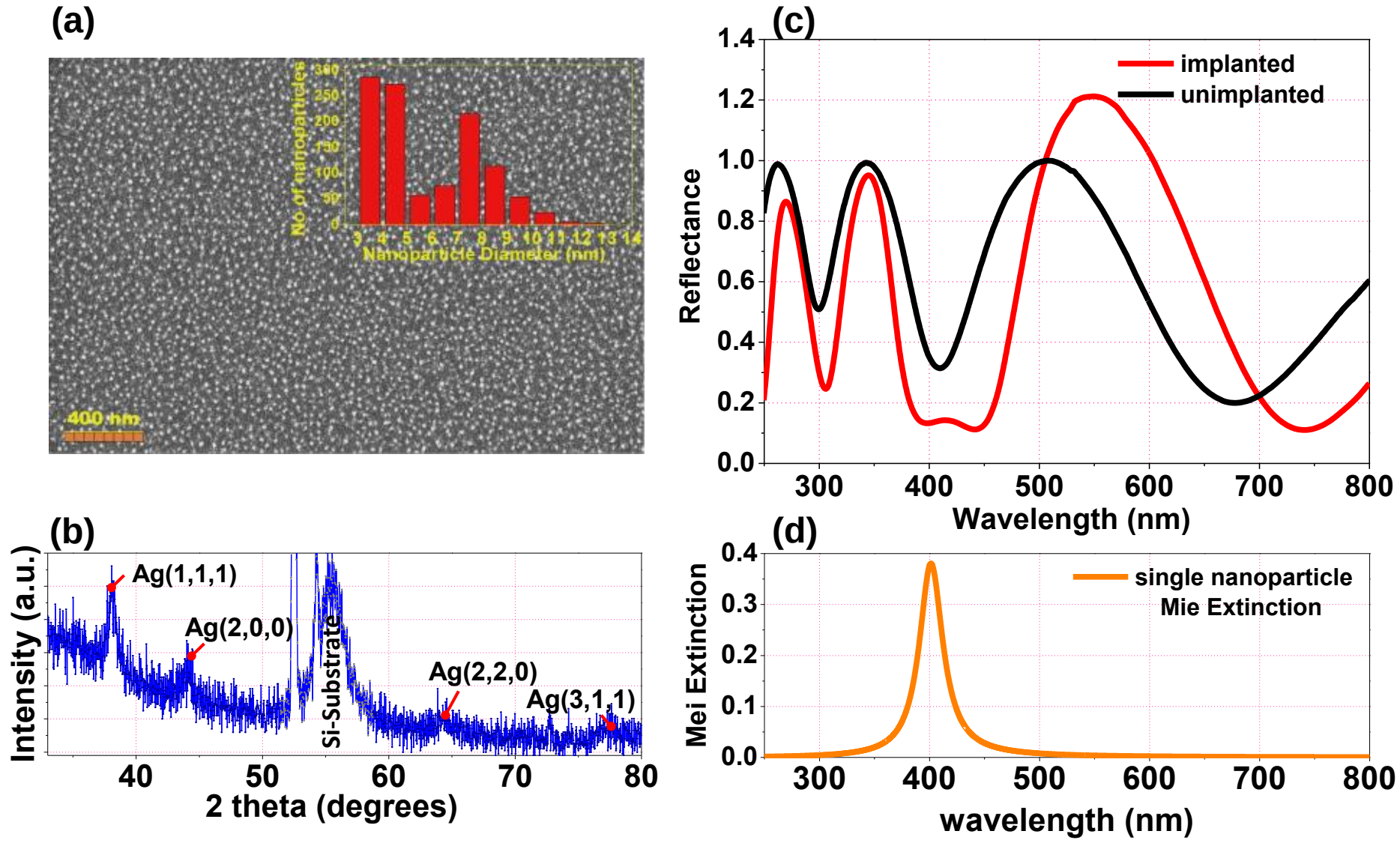

FIG.2.(a) The surface morphology of the Ag implanted thermal oxide surface as observed by SEM. Each division in the image scale represents a length of 50nm. (b) The GIXRD for the Silver implanted thermal oxide. The broad peak around 55º corresponds to the signal from the Si substrate. (c) Reflectance spectra for the implanted (blue) and the un-implanted (red) films. (d) The optical Mie extinction spectra for a single nanoparticle calculated as per the Mie theory[11].

The XTEM micrograph of the Ag ion-implanted $SiO_2$ thin film is depicted in figure 1a. The formation of the crystalline Ag NPs in the as implanted films can be attributed to ion beam induced local heating, as evidenced and explained in previous reports[8]. Further, the formed NPs show a wide distribution of radii ranging between 2 to 12 nm (figure 1b). The smallest particles are found at a certain depth in the film, whereas moving towards the surface the particle size increases. The largest particles are found to protrude out of the film surface indicating the existence of an outward diffusion[8] of the Ag ions. Formation of the particles on the film surface is confirmed by SEM, showing the formation of particles with of sizes ~20nm and less, as depicted in figure 2a. The formation of crystalline Ag NPs is also confirmed by GIXRD (figure 2b). LSPR properties of the Ag NP film is investigated with UV-Vis reflectance spectra. The comparison of the optical reflectance of the Ag NP film with the un-implanted film is depicted in figure 2c. Multiple oscillations are due to thickness dependent interference, whose wavelength minima ($\lambda_m$) depend on the optical path length traversed by a certain wavelength of light. This is expressed by the condition:

$$\lambda_m = (2nd\cos\theta_i)/(m + \tfrac{1}{2}) \qquad (1)$$

Here 'm' is a positive integer representing the order of the interference minima, $\theta_i$ is the angle of incidence, 'd' being the film thickness and 'n' is the refractive index of the film. Coming back to figure 2c, it must be noted that two dips in the reflectance in the range 400 nm to 450 nm cannot be accounted for by the reflectance minima described under equation 1. Therefore one of these dips originates from the LSPR related absorption in the silver NPs present in the implanted sample.

Also plotted in the graph is the Mei optical extinction, for 20 nm Ag NPs embedded in $SiO_2$, in agreement with C*arles et al.*[11]. It must be noted that optical extinction accounts for the combined effect of scattering as well as absorption. For these aforementioned calculations, the Fermi length for electrons and the plasmon frequency in silver have been taken to be 54.7 nm and $1.39\times10^{16}$ rad/sec, respectively. From the extinction peak (centered at 400 nm) it may appear that out of the two dips in the range 400-450 nm the dip in the reflectance at 400 nm originates from LSPR related absorption. However, we cannot with certainty associate the peak at 400 nm with LSPR due to silver nanoparticles, since in the present case the particles are not entirely embedded in the $SiO_2$ matrix, rather there is a high density of silver nanoparticles protrusions on the surface of the $SiO_2$. In fact, in recently carried out work by *Liu et al.*[12] it has been found that the thickness of the dielectric layer beneath the Ag NPs also plays a role in determining the position of the LSPR resonance absorption. Further, the calculation of the optical extinction (red curve in figure 2d) does not take into account the size distribution of NPs as well as it neglects the higher order multipole terms given by Mei scattering theory.

It is understood now from the above discussion that the reflectance from $SiO_2$ film containing the silver NPs is a convoluted mixture of the LSPR related absorption from silver NPs and the thin film interference in the $SiO_2$ thin film. It means that the experimentally obtained reflectance spectral feature (figure 2c) for the implanted sample cannot be obtained from a simple combination of the un-implanted film and the calculated Mei Extinction of the Ag NPs (figure 2c). The aforesaid conclusion can be attributed to two reasons, firstly, the closely spaced nanoparticles embedded in the medium no longer behaves as isolated nanoparticles in a medium rather the incident light sees the implanted film as being composed of several layers. Wherein, the layer containing the embedded Ag NPs, behaves as composite medium, who's dielectric function have to be determined from an effective medium theory. Secondly, introduction of the ions into the dielectric matrix can also lead to change in the net thickness of the films through various mechanisms[8]. Further, there is a possibility of the coupling of the thin film interference modes with the LSPR in NPs[11]. The extent of this coupling would depend on the field intensities arising out of the thin film modes at the depth at which the NPs are located. Thus the transfer matrix theory[13] for optical simulations was employed in combination with the effective medium approximation, to simulate the reflectance arising from the present composite thin film. The film structure was assumed to be composed of 4 layers, as shown in figure 3a, to describe the combined effects of differently sized particles in air or $SiO_2$ medium. Details of the formulae used to calculate the effective dielectric constants for an ensemble of NPs embedded in $SiO_2$[14] within close proximity to each other has been imbibed from *Garcia et al.*[15]and *Hovel et al.*[16], wherein, the average size (>30 nm) of the nanoparticles is well below the incident wavelengths (200-800nm). The details of the calculation of effective dielectric constants and subsequently, the effective n and k are provided in the supplementary section. The effective n and k calculated in the aforementioned way for an ensemble of 20nm Ag NPs embedded in SiO2 (corresponding to layer-2) are plotted in figure 3b and 3c, respectively. It must be noted here that the extinction spectra obtained for an ensemble of NPs (figure 3b) is different in spectral shape, when compared to the single NP extinction displayed in figure 2d. Coming back to figure 3a, the layers identified within the implanted films are described here. The first three layers from the film surface are composite layers. Layer-1 was modeled as 20 nm Ag NPs in air medium with 10% volume fraction, followed by a layer-2 composed of 20 nm Ag NPs in $SiO_2$with 10% volume fraction, and layer-3 has 5 nm Ag NPs with 4% volume fraction in $SiO_2$. The layer-4 is a layer of pristine $SiO_2$. The effective n & k in figure 3b & 3c thus correspond to the layer-2. Similar to the calculations for the other layers depicted in figure 3a were also determined and are provided in the supplementary file. The obtained optical constants of the layers together with the thicknesses of these layers are modeled into the transfer matrix theory to obtain the reflectance from the stratified system. The simulated reflectance spectrum and its comparison with the normal incidence reflectance spectrum are depicted in figure 3d. It may be observed here that the transfer matrix (TM) -simulations qualitatively reproduce the double kink in the vicinity of 400 nm. However, the qualitative differences can be attributed to the assumptions made in modeling the reflectance. Namely, the assumed mono-dispersity for the given layers, and further non-specular reflection might also cause the difference in the amplitudes of the interference oscillations observed between the experiment and calculation. The transfer matrix simulations also allow us to predict the spectra for different angles of incidence. Therefore, for a comparison of theory and experiment, optical reflectance was also recorded at different angles of incidence, ranging between 0º-45º in steps of 5º. The experimental curves so obtained are compared with the corresponding simulated reflectance obtained from Transfer matrix (TM) simulations, as depicted in figure 4.These angle dependent measurements clearly show the coupling of the LSPR and the bilayer interference is maximum at normal incidence however at higher incidence angles the strength of the LSPR related absorption weakens. Unlike the experimental observations made above, if the coupling effect were absent, the LSPR related dip would have been independent of the angle of incidence (both in strength and peak position), while the interference minima shift left because of an increase in the optical path at higher angles. The strong LSPR observed in the normal incidence reflectance of the implanted $SiO_2$ film therefore results from a positive coupling of the interference modes with the plasmon resonance in the embedded silver nanoparticles. The transfer matrix simulations have further been used to calculate the field intensity profiles through the depth of the films using a methodology adopted by *Ohtaet al.*[17]. For incident un-polarized light the net field intensity as a function of film depth is defined as:

$$F(z) = \left[F_x(z) + F_y(z) + F_z(z)\right]/2 \tag{2}$$

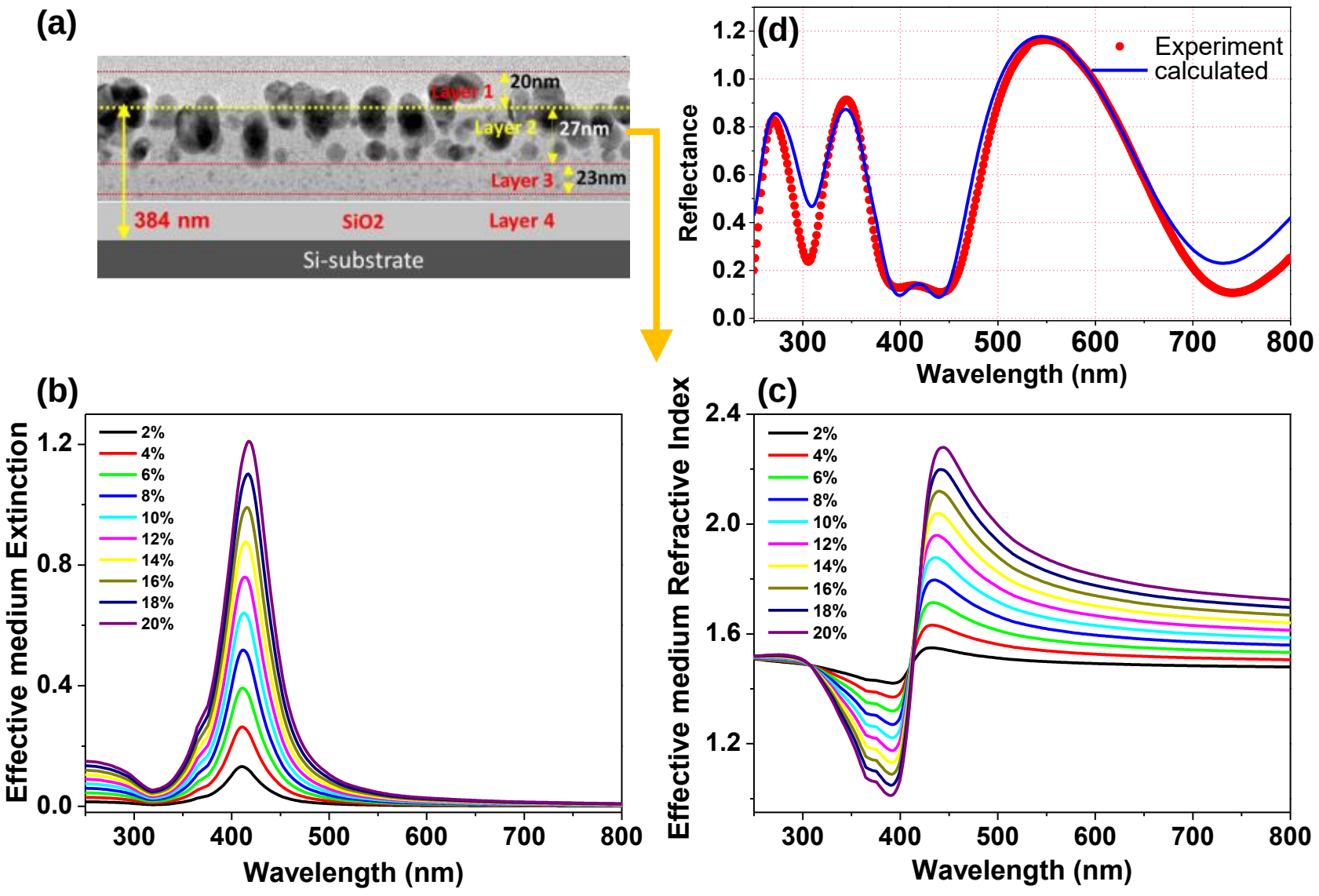

FIG.3. (a) The schematic of layers 1 to 4 (not to scale) identified within the Ag implanted film. The yellow dotted line corresponds to the film surface. Effective refractive index (b) and extinction (c) for an ensemble of Ag NPs (20 nm) embedded in $SiO_2$ matrix (Layer-2), for various volume fractions of the NPs inclusions. (d) The simulated fit of the normal incidence reflectance from the transfer matrix approach using the effective optical constants of the various layers assumed in figure (a) the red dots represent the experimentally obtained data.

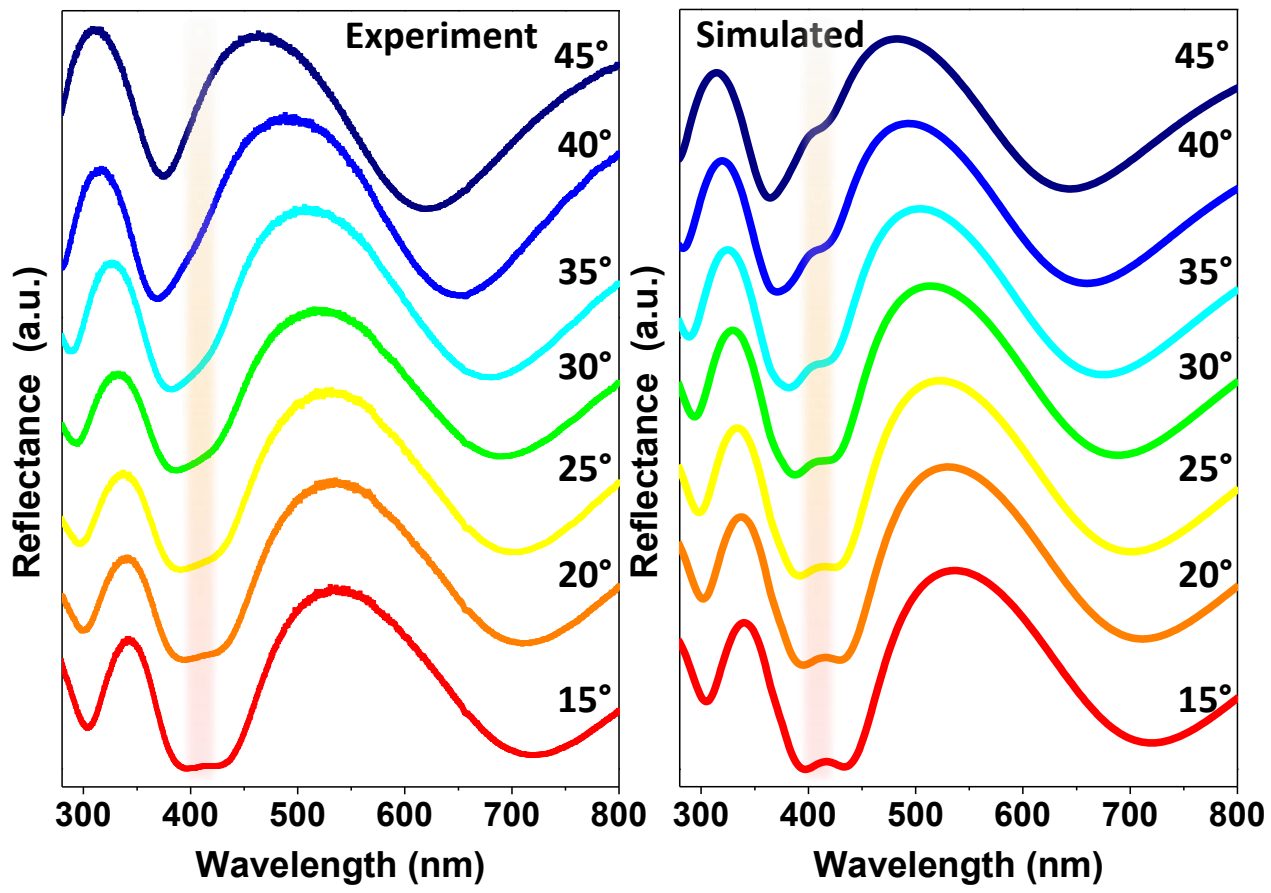

FIG. 4. (a) Experimentally obtained Angle dependent Reflectance spectra for a stratified structure depicted in figure 3a (b) transfer matrix simulations of the same. The shaded area in the background shows the LSPR active region for Ag NPs implanted in $SiO_2$.

Here, $F_x$, and $F_y$ correspond to net field intensities parallel to the plane of the film, wherein $F_x$ and $F_y$ lie parallel and perpendicular to the plane of incidence. The details of the approach are provided in the supplementary section. Figure 5a and 5b show the calculated net field intensity profiles, averaged over wavelengths between 200 nm-800 nm for un-implanted and implanted $SiO_2$ thin films for different incidence angles. The layers identified in figure 5a (implanted film) are the same as used in figure 3a. The field profiles very clearly depict the strongly enhanced fields at the film surface for the implanted films (figure 5a) in sharp contrast to the un-implanted film (figure 5b). Also one can notice the decrease in the field within the films and its concomitant increase at the film surface with increasing angles of incidence. The decreased field intensities within the NP containing layers, layer-1 & layer-2 can also be noticed and is understood to be due to attenuation of the electromagnetic field in the Ag NP dielectric composite layer. The enhanced electric field at the film surface clearly underlines the use of these structures as SERS substrates.

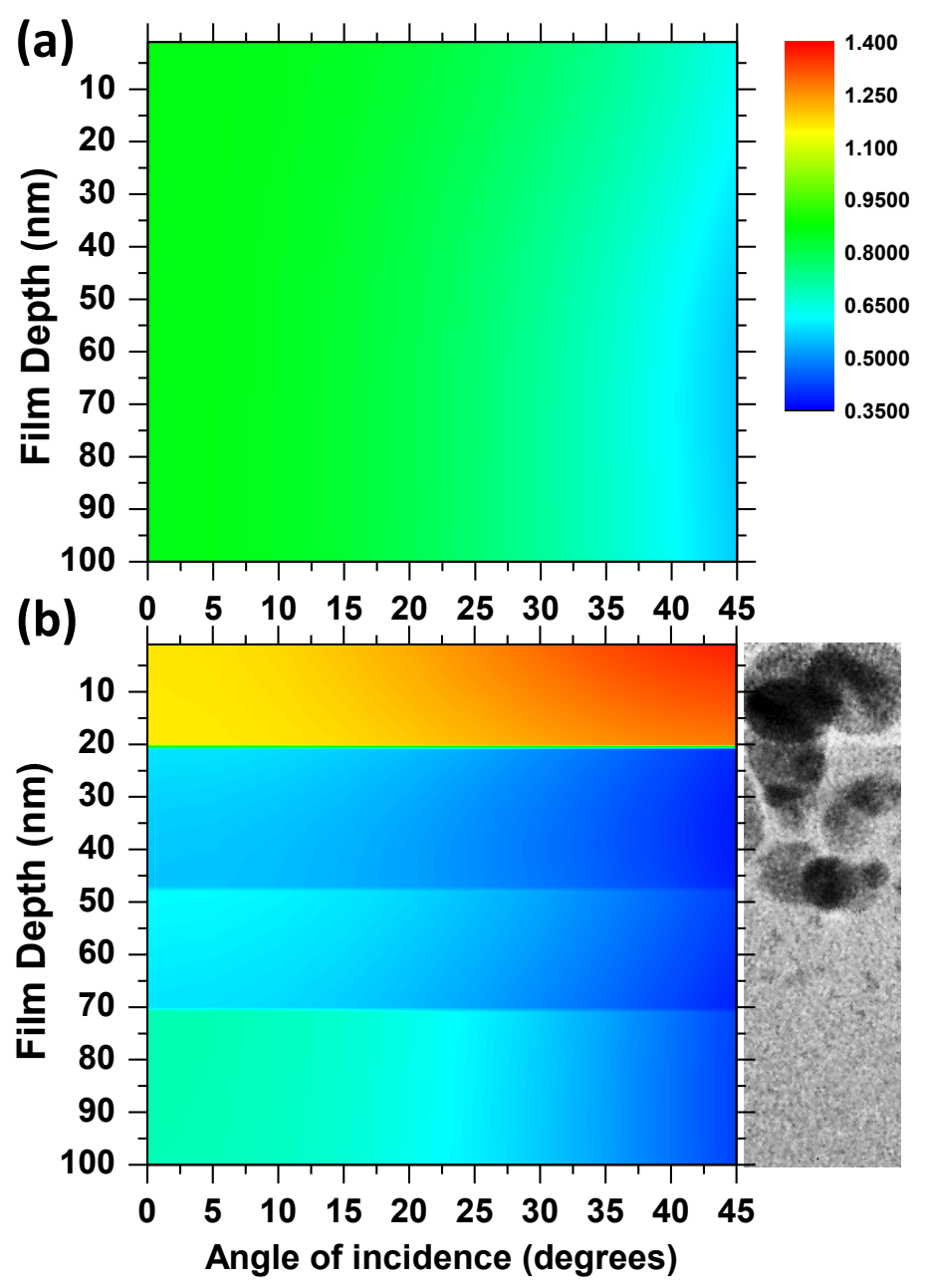

FIG.5. The simulated field intensity depth profiles for (a) un-implanted film and (b) an implanted $SiO_2$ film. The representation for various implanted layers is the same as used in figure 3a. The color scale on the right applies to both the color maps.

In summary, low energy silver negative ion implantation in thermally grown $SiO_2$ was carried out with the aim of growing embedded NPs. An outward diffusion of silver NPs from the film surface is observed, and is attributed to the ion beam induced heating. The UV-Vis reflectance spectrum of the implanted film shows a double kink feature close 400nm. We have attempted to identify the kink arising out of LSPR in the Ag NPs. It is found that, TM simulations in combination with the standard effective medium approximations are found to accurately reproduce the LSPR related features in the normal incidence reflection spectra of Ag implanted $SiO_2$ thin film. The TM simulations have been further used to estimate the field enhancements at the film surface arising from the coupling of the thin interference modes with the LSPR in metal nanoparticles. These latter simulations indicate the presence of strong electric fields at the film surface, which in turn would be immensely useful in designing various applications.

Institute of Physics Bhubaneswar and DST-SERB are acknowledged for the National Post-Doctoral Fellowship (grant no. 2016/PDF/002115). XRD facility of IIT Delhi is acknowledged for the XRD measurements. Inter University Accelerator Center New Delhi and its personnel at the low energy ion beam facility are acknowledged for the implantation experiments. Prof. P. V. Satyam and Dr. Arun Thirumurugan are acknowledged for the SEM and XTEM measurements.

## SUPPLEMENTARY MATERIAL

**Calculation of the effective n & k for composite layers**:

For the calculation of the effective dielectric function of a given layer, firstly the modified dielectric function of Ag NPs was determined. This involved the modification of electron collision frequency as imposed by the NP boundaries. The modified collision frequency for NPs is given by:

$$\gamma' = (Av_F/L) + (Av_F/R) = (\gamma_0 + Av_F/R) \quad (2)$$

here$\gamma_0 = Av_F/L$ represents the collision frequency for bulk silver, $v_F$ is the Fermi velocity for silver and taken as $1.4\times10^6$ m/s, L is the electron mean free path for silver and taken as 57nm, $R$ is the NP radius, $A$ is the NP size parameter and taken as 1 in the present case. The above modified collision frequency is the used to describe the correction term to be added to the bulk silver dielectric functions, in order to obtain the dielectric constant of the NPs. The expression to obtain the NP dielectric function from the bulk was given by *Hovel et al.*[15] as:

$$\varepsilon(\omega, R) = \varepsilon_{bulk}(\omega) + \frac{\omega_p^2}{\omega^2 + i\omega\gamma_0} - \frac{\omega_p^2}{\omega^2 + i\omega(\gamma_0 + Av_F/R)} \quad (3)$$

Here, $\gamma_0$ is the bulk silver collision frequency, $\omega_p$ is the bulk plasmon angular frequency for silver whose value is taken to be $1.4\times10^{16}$ rad/s. $\omega$ is the angular frequency of the incident light. The definitions of $R$ and $A$ are as mentioned above. Once the NP dielectric function was determined in this way, the NP plus dielectric effective medium dielectric constant is obtained from the formulae elaborated by *Garcia et al.*, wherein, the real and imaginary parts of the effective dielectric function $\varepsilon_{ef}$ is given by:

$$\varepsilon_{ef1} = \varepsilon_m + \frac{ac+bd}{c^2+d^2} \quad (4a)$$

$$\varepsilon_{ef2} = \frac{bc-ad}{c^2+d^2} \quad (4b)$$

Where, parameters a, b, c and d are defined as follows:

$$a = f(\varepsilon_1 - \varepsilon_m)\ ,\ b = f\varepsilon_2 \quad (4c)$$

$$c = \varepsilon_m + \beta(\varepsilon_1 - \varepsilon_m) - fg(\varepsilon_1 - \varepsilon_m) \quad (4d)$$

$$d = \beta\varepsilon_2 - fg\varepsilon_2 \quad (4e)$$

Here $f$ is the volume fraction of embedded Ag NPs, $\varepsilon_m$ represents the dielectric constant of the medium, $\beta$ includes the effect of NP shape and taken as 1/3 for spherical NPs. Lastly, the parameter $g$describes the interaction of NPs in close proximity of each other and is defined as:

$$g = \frac{1}{3\varepsilon_m} + \frac{K}{4\pi\varepsilon_m} \quad (4f)$$

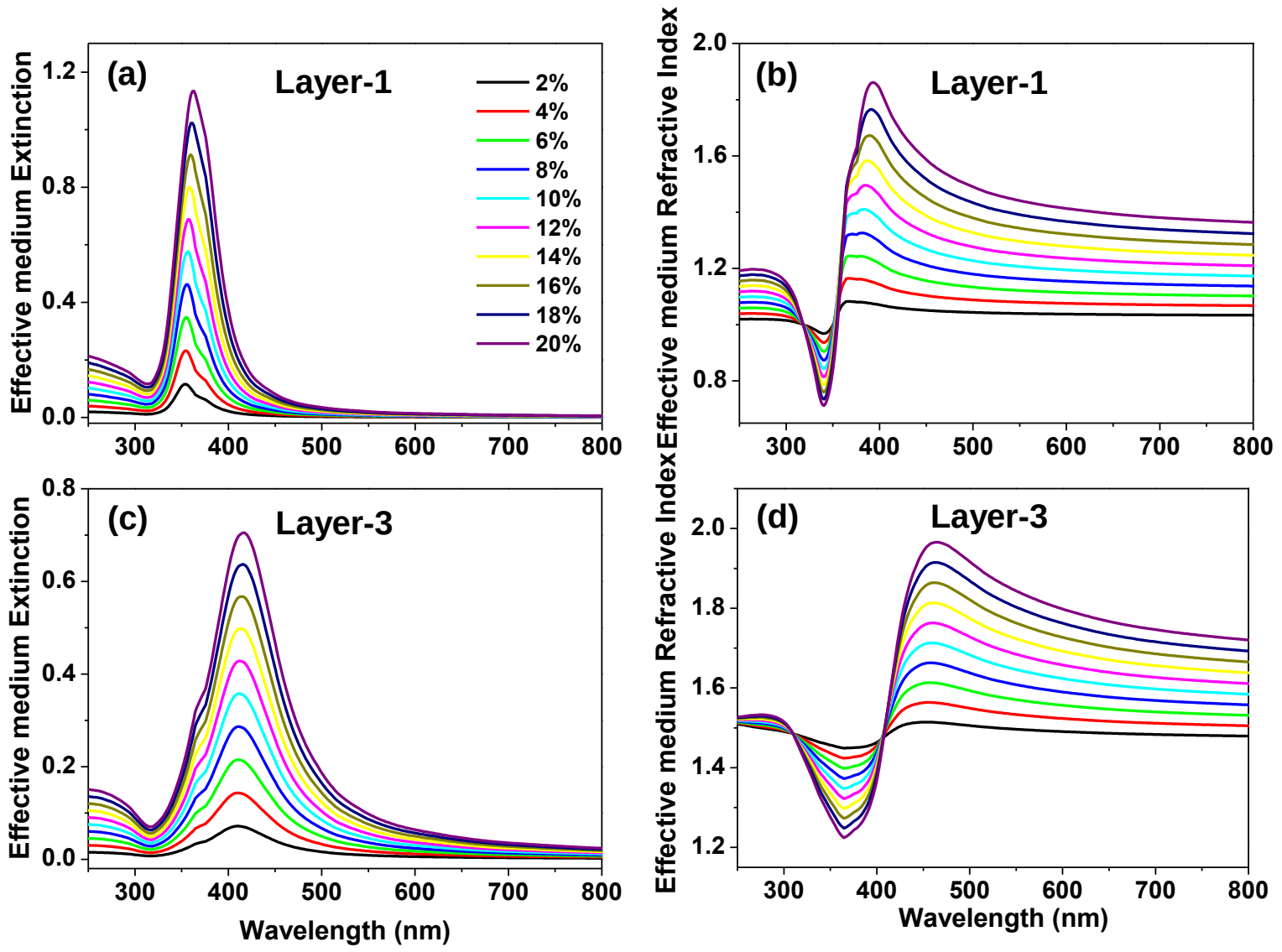

FIG.6. Effective (a) refractive index and (b) extinction for 15nm Ng NPs in air medium (layer 1). Effective (c) refractive index and (d) extinction for 5 nm NPs embedded in $SiO_2$ (layer 3).

Here, $K$ is the dipolar interaction parameter, wherein zero value corresponds to the absence or cancelling out of dipolar interaction. A positive value occurs when the dipole interaction reinforces the applied field at the location of the NPs, whereas; a negative $K$ value represents the converse. The effective refractive index and extinction for each layer is determined from the real and imaginary parts of the effective dielectric constant in equations 4a and 4b.

**Calculation of the field intensity at a particular depth of the film**:

These calculations, employ the transfer matrix method, to carryout superposition of various transmitted (+) and reflected (-) fields at a given layer depth and layer of the film. The fields for the s, p polarizations are determined and appropriately superposed to get the $E_x$, $E_y$ and $E_z$ components of the net electric field at a certain film depth.

$$E_x(z) = \left[E_p^+(z) - E_p^-(z)\right] \cos\theta_j \quad (5a)$$

$$E_y(z) = \left[E_s^+(z) + E_s^-(z)\right] \quad (5b)$$

$$E_z(z) = \left[E_p^+(z) + E_p^-(z)\right] \sin\theta_j \quad (5c)$$

Where $\theta_j$ is the angle of refraction in the given layer j. $E^+$ and $E^-$ correspond to waves along and opposite to the incident light, whereas the subscripts s and p indicate the perpendicular and parallel electric field components.

Corresponding Field intensity ratios are defined as:

$$F_x(z) = |E_x(z)|^2/\left|E_{op}^+\right|^2 \quad (6a)$$

$$F_y(z) = \left|E_y(z)\right|^2 / |E_{os}^+|^2 \tag{6b}$$

$$F_z(z) = |E_z(z)|^2 / \left|E_{op}^+\right|^2 \tag{6c}$$

Here, $E_{os}^+$ and $E_{op}^+$ correspond to the perpendicular and parallel Electric field components of the incident wave. The net field intensity ratio at a depth z, for the case of unpolarised incident light is given by:

$$F(z) = \left[F_x\,(z) + F_y\,(z) + F_z\,(z)\right]/2 \tag{7}$$